# Implicit and Explicit Research Quality Score Probabilities from ChatGPT

Mike Thelwall, School of Information, Journalism and Communication, University of Sheffield, UK. https://orcid.org/0000-0001-6065-205X

Yunhan Yang, Faculty of Education, The University of Hong Kong, Hong Kong SAR, China. https://orcid.org/0000-0003-0210-7285

**Abstract:** The large language model (LLM) ChatGPT's quality scores for journal articles correlate more strongly with human judgements than some citation-based indicators in most fields. Averaging multiple ChatGPT scores improves the results, apparently leveraging its internal probability model. To leverage these probabilities, this article tests two novel strategies: requesting percentage likelihoods for scores and extracting the probabilities of alternative tokens in the responses. The probability estimates were then used to calculate weighted average scores. Both strategies were evaluated with five iterations of ChatGPT 4o-mini on 96,800 articles submitted to the UK Research Excellence Framework (REF) 2021, using departmental average REF2021 quality scores as a proxy for article quality. The data was analysed separately for each of the 34 field-based REF Units of Assessment. For the first strategy, explicit requests for tables of score percentage likelihoods substantially decreased the value of the scores (lower correlation with the proxy quality indicator). In contrast, weighed averages of score token probabilities slightly increased the correlation with the quality proxy indicator and these probabilities reasonably accurately reflected ChatGPT's outputs. The token probability approach is therefore the most accurate method for ranking articles by research quality as well as being cheaper than comparable ChatGPT strategies.

## 1   Introduction

Research quality evaluation for periodic departmental reviews, appointments, promotion and tenure is a common and time-consuming expert review task. The use of automated or quantitative methods to support, replace, or verify human judgements about the quality of academic research is therefore an important need for research evaluators and managers as well as practicing researchers. There are many reasons for using indicators in this context, including the unavailability or high cost of expert judgements at the scale needed for research managers/evaluators, a mistrust in subjective judgments, and a need for objective validation, perhaps a special concern for quantitative researchers.

### 1.1   Citation-based indicators and research quality

In response to the need for methods to support expert judgement in research evaluation, citation-based indicators like the Journal Impact Factor, career citations for academics, citation counts for individual articles, and field and year normalised citation rates have become widely adopted and sometimes even regarded as gold standards for research quality (Chen & Lin, 2018; Minoura et al., 2024; Robinson-Garcia et al., 2023; Rushforth & Hammarfelt, 2023). For example, researchers reported that their work was assessed entirely (16%), mostly (39%), equally (28%, with qualitative) or partly (10%) with indicators according to a SpringerNature 2024 international multidisciplinary survey of 3,827 researchers, with only 2% reporting exclusively qualitative assessments (Farr, 2025; SpringerNature, 2025).

Despite the importance attached to citation-based indicators, they have many limitations and need to be used cautiously (Hicks et al., 2015). Citations can be negative (to criticise the cited work) or trivial for background information. Their selection can be biased in favour of the work of colleagues or well-known academics, and important citations can be omitted as too obvious to need mentioning (MacRoberts & MacRoberts, 2018; McCain, 2011). More fundamentally, work can have value without being cited, such as when it disproves previous studies, shutting down a line of enquiry, or when it leads to practical applications, such as a medical cure, a product innovation, or improvements in professional practice. For these reasons, when groups of academics try to define research quality, they tend to avoid citations but usually focus on three core dimensions: rigour, originality and significance (Langfeldt et al., 2020).

Here, rigour is the extent to which the findings are reliable and that the reported research has been conducted with sufficient care to reduce the chance of false conclusions. Although no non-mathematical study can claim absolute proof, studies vary in the extent to which the chances of false conclusions are reduced (e.g., with larger or more varied sample sizes). In contrast, significance or impact relates to the actual or likely future scholarly or societal value of the work. Citations might be taken as an indicator of scholarly significance, although with the limitations noted above (Aksnes et al., 2019). Finally, originality relates to the extent to which the research goal, methods, or findings are novel in any way (Langfeldt et al., 2020).

## 1.2 Large language models (LLMs) and research quality

Large language models have provided the first realistic possibility to automate quality scoring in terms of directly assessing originality, significance and rigour. This is possible because they can explicitly be asked to evaluate these characteristics, an immediate advantage over citations, which may be primarily restricted to scholarly significance (Aksnes et al., 2019). Previous studies have shown that with a prompt requesting a research quality score, ChatGPT 4o and 4o-mini tend to give results that correlate more highly with an independent indicator of research quality (departmental average quality scores) than long term citation rates in most fields (Thelwall, 2025c; Thelwall & Yaghi, 2024; Thelwall et al., 2025). Google Gemini has a similar, but weaker, ability (Thelwall, 2025b). LLMs have an additional advantage over citations in that they can be applied to just published work, without a delay of several years, as needed for an adequate citation window (e.g., Wang, 2013). This is useful for research evaluation purposes, for which older research is less relevant, or even unavailable for early career scholars.

LLM quality scores improve when averaged over multiple repetitions. In other words, if ChatGPT is asked to score an article 30 times then the average of these 30 scores is a better indicator (has a higher correlation with expert scores) than each individual score. A possible explanation for this is that averaging repetitions gives information about ChatGPT's certainty for its scores. For example, if it returns 30 scores of 3* then this reflects more certainty about the 3* score than if it returns 25, with five 4* scores. Moreover, the latter case suggests that the article might be better than the average 3* article. In this example, as in most cases tested so far, the UK Research Excellence Framework (REF) quality scale has been used and this consists of four integers: 1* (nationally relevant), 2* (internationally relevant), 3* (internationally excellent), and 4* (world leading). It seems obvious that these scores are simplifications and that most articles would score somewhere between these values (e.g., 3.5*), so averaging multiple ChatGPT scores might even, in theory, start to reflect the "true" score of some articles. Of course, research quality is a disputed human construct for which there may not be a genuinely "true score" in the sense of a value that almost all experts would agree on.

Because of the advantage of property, previous studies have queried ChatGPT between 5 and 30 times to get average scores (Thelwall & Yaghi, 2024). This multiplies the financial cost of the information since each API query must be paid for. In addition, the result is only an approximation to the underlying probability model, no matter how many queries are submitted. Thus, it is logical to seek an alternative approach that would get more quickly or directly to the underlying ChatGPT probabilities for the different scores. This assumes that ChatGPT's training gives it the ability to reveal, through multiple queries, the extent of its knowledge about the possible correct scores, and that this is separate from its ability to know the correct score (e.g., Ahdritz et al., 2024).

### 1.3 From single LLM scores to score probability distributions

This article introduces two novel approaches for article quality scoring: **classification percentage requests** and **token probability leveraging**. Here a classification percentage request is defined to be a ChatGPT prompt asking for the likelihood (in percentages) that the submitted article would be assigned each of the four scores on the scale. This is an explicit request to ChatGPT to estimate the probabilities that it would report the different possible scores. The results might reveal, for example, that its best guess was 3* but it considered that there was also a 40% chance of 4*. This might be equivalent to submitting the original prompt 100 times and getting 3* scores 60 times and 4* scores 40 times.

In contrast, the token probability leveraging approach is to request that ChatGPT gives a single score but also extracting information from it about the probabilities of the tokens in its response and when it gives the score (e.g., 3*) and using those probabilities to identify the probability that it could have given a different score (e.g., 2*) instead. This gives direct access to the underlying probability model, at least as formulated for the response. Of course, for most LLMs the probability model is influenced by random factors as well as initial outputs and so varies for identical prompts.

This article assesses whether the two approaches above give improved results compared to the standard approach previously used of requesting a single score and averaging the scores across multiple iterations. Previous attempts to vary ChatGPT's prompts and settings (e.g., temperature) to improve the results have been unsuccessful (Thelwall, 2025b), but score probability distributions have not been tried before. The research questions below will be addressed through correlation tests.

- RQ1: Do **classification percentage requests** give more useful research quality scores from ChatGPT than standard prompts?
- RQ2: Does **token probability leveraging** give more useful results than accepting the single recommended ChatGPT score?
- RQ3: Are the probabilities from the two sources internally consistent, in the sense of accurately predicting the distribution of scores from other ChatGPT sessions?

## 2   LLMs for classification

Background information about LLMs is given here by reviewing findings about zero-shot text classification. The default type of LLM for this is the bidirectionally trained Encoder type, like BERT (Devlin et al., 2018), which is optimised for interpreting words in their context. In contrast, Generative LLMs like ChatGPT (Brown et al., 2020) are Decoder-only, trained to predict the next word (more specifically, token: a short sequence of characters) based on the prior context. There are also Encoder-Decoder models, like BART (Lewis et al., 2019). The most powerful LLMs currently seem to be the Decoder-only type. Instruction-tuned models have their text generation capacity adapted to follow instructions by being fed with many examples of instructions and correct responses (Zhang et al., 2023). A large investigation found that seven Encoder-only or Encoder-Decoder LLMs with between 110 million and 11 billion parameters outperform previous approaches to detect text topics in a set of 23 public datasets, with fine tuning helping and the larger instruction-tuned models performing best (Gretz et al., 2023).

### 2.1.1   Zeros-shot direct classification

This section discusses various zero-shot (i.e., no training data and no examples) classification tasks with LLMs for which different approaches have been compared. Whilst the published research suggests that LLM-based approaches tend to outperform others there is not a consensus about when zero-shot, few-shot, fine-tuning work best, which prompting strategy is optimal, and which type of LLM works best (e.g., Wu et al., 2025). This may be because the error margins are relatively small, so one LLM may perform well on a task compared to others not because it has a greater underlying capability but because it was "lucky" in the sample selection – with other LLM performances occurring within 95% confidence limits (if reported). For example, a review of LLMs in healthcare text classification did not make general recommendations about the best strategy (Sakai & Lam, 2025). The issue is complicated by ChatGPT's good performance on a wide range of text processing tasks (Kocoń et al., 2023) and the frequent release of new versions of it and its competitors. Some text classification task performances are reported here to illustrate the variety of findings.

A test of ChatGPT 4 (not clear which version), the downloadable Decoder-only LLM Qwen2.5-7B and others for four datasets, with testing on up to 2,974 items suggested that ChatGPT did not need fine tuning if there were few categories (e.g., three), but this conclusion would need to be verified on other datasets (Vajjala & Shimangaud, 2025). In contrast, another study found that fine tuning allows smaller LLMs to outperform larger LLMs like ChatGPT on a task with few categories (Bucher & Martini, 2024). Thus, the optimal approach may vary by task. Conversely again, ChatGPT 4o-mini had similar performance to a fine-tuned ChatGPT 4o for a sentiment analysis task (Beno, 2024), and ChatGPT 4o-mini outperformed smaller LLMs after fine tuning on another task (Roumeliotis et al., 2024).

### 2.1.2   Classification percentages

LLMs can be asked to report their certainty about their responses to prompts (Lin et al., 2022; Xiong et al., 2023) and there have been claims that the larger models have a good ability to assess their own level of confidence in their answers (Kadavath et al., 2022). Thus, if they are asked how certain they are about an answer then their reply may tend to be reasonable. A comparison of different LLMs has found that smaller and weaker models tend to be more optimistic and less realistic when asked to self-report their confidence in an answer (Omar et al., 2025). ChatGPT 4o should therefore be a top performing model in this regard.

This ability to accurately self-report answer certainty has been questioned for ChatGPT (Gallifant et al., 2024; see also: Hurst et al., 2024). For example, the self-reported confidence claims of ChatGPT 4o have been found to be poor for a diverse set of questions (Pawitan & Holmes, 2024). Moreover, if chain-of-thought type prompts are used to ask ChatGPT to justify an answer then this increases its self-reported confidence in the answer, irrespective of whether it is correct (Fu et al., 2025). A comparison of human and ChatGPT 4o answer confidence found that ChatGPT consistently overestimated the likelihood that its answer was correct, whereas the humans tested tended to be more realistic (Sun et al., 2025). Thus, ChatGPT's self-reported answer confidence ability seems to be weak.

No previous study seems to have requested confidence estimates for multiple alternative answers. This might be thought of as a multi-task problem: making multiple simultaneous estimates. Counterintuitively, ChatGPT seems to answer slightly more accurately when given multiple tasks in a single prompt (Son et al., 2024), so it might be better at creating tables of confidence estimates than at creating single estimates. In support of this, asking for a table of estimates might entail evaluating each option separately and then comparing the answers.

### 2.1.3 Token probabilities

The relative likelihood of a finite set of answers from an LLM can be obtained by submitting the prompt and then obtaining the probabilities of all the possible answers. These probabilities can be obtained by extracting the log probabilities of the tokens representing the answers as potential next output tokens for the LLM (Duan et al., 2024; Wang et al., 2024; Zhao et al., 2024). This works for questions where the answer is a single token, such as in multiple choice questions, but not when the answer is longer, such as an explanation.

Not all LLMs expose token probabilities (e.g., Gemini does not at the time of writing) and ChatGPT only exposes a maximum of five tokens with the highest probabilities (by including "logprobs":true,"top_logprobs":5 in the request). When using this approach, it should not be assumed that the first token returned is the answer because it might instead be a preliminary comment (Wang et al., 2024). Identifying token probabilities may give a good approximation of the frequencies with which the different tokens would appear as answers, given multiple identical queries (Ahdritz et al., 2024).

A few previous studies have exploited token probabilities, although none in the same way as here. An investigation of the answer correctness probabilities reported by a range of offline LLMs (e.g., Meta-Llama-3.1-70B-Instruct) on request (i.e., the prompt asks for the answer and the percentage/probability that it is correct) with token probabilities extracted from the answer tokens and potential alternative answer tokens that were not used. It found that the implicit probabilities from tokens were more accurate than explicit probabilities given in answers to health questions, especially for the smaller models (Gu et al., 2024). Token probabilities have also been used to help detect hallucination in answers (Quevedo et al., 2024). Finally, an evaluation of a more complex probability-based approach (not direct weighting) for open weight (offline) LLMs found it to give small improvements on a range of tasks (Duan et al., 2024). The direct weighting approach for numerical outputs, as in the current paper, does not seem to have been attempted before.

## 3 Methods

The research design was to calculate quality scores for a large set of articles from multiple fields with both **classification percentage requests** and **token probability leveraging** and then to assess whether the results of either correlate more strongly with research quality scores than scores from the previously used approach. The prompts, as described below, were submitted to ChatGPT 4o-mini (4o-mini-2024-07-18, for compatibility with data previously reported) between 6 and 13 April 2025.

### 3.1 Data: Articles and expert research quality scores

The dataset is the same as previously used, REF2021 journal articles without short abstracts and classified into the 34 REF Units of Assessment (UoAs), each of which is a field or group of fields. Here a "short abstract" is one that is in the shortest 10% for a UoA. This restriction excludes articles without abstracts, articles with accidentally truncated abstracts, and a higher proportion of short form articles that are not fully equivalent to most articles. Abstracts are important because the ChatGPT scores are obtained from titles and abstracts alone.

The same dataset as before was chosen for two reasons. First, it is the only large science wide set of articles with research quality evidence, so there is no alternative for a science-wide study, other than citations, which do not reflect all aspects of research quality. Second, using the same dataset allows the results to be directly compared.

The dataset consists of standard journal articles, excluding reviews, that were first published between 2014 and 2020 and that were selected by UK universities to be entered into the national REF2021 evaluations. These are therefore self-selected to be the highest quality UK research, subject to the following main conditions: (a) at least one author had to be working at the UK institution on the REF census date in 2020, (b) each author could submit between 1 and 5 outputs, with other types like books also eligible, and (c) the department had to submit 2.5 outputs per full time equivalent researcher, on average (REF2021, 2019). Thus, the dataset is almost exclusively UK-authored (at least partly) and with above average quality for the UK.

Quality scores were assigned to each article by one of the 34 sub-panels of experts (usually professors or other senior researchers) using the four-point scale mentioned above. There is one sub-panel for each UoA, and these are either broad fields (e.g., UoA 8 is Chemistry) or groups of related fields (e.g., UoA 24 is Sport and Exercise Sciences, Leisure and Tourism). The assessors were guided by training and published quality criteria (Wilsdon et al., 2015), and the evaluation process took a year. As a matter of policy, although each individual output is given a score, these scores have been destroyed and only the percentage of outputs scoring 1*, 2*, 3* or 4* was published for each submission. Here, a "submission" usually means all outputs submitted by a single higher education institution (out of the 157 assessed) to a single UoA (out of 34). For intuitive simplicity a submission is described as a "department" in the current article because it combines a field and institution focus, even though many and possibly most submissions do not equate to departments because other organisational names are used (e.g. school) and submissions might include research from multiple departments or only one section within a department.

Since individual article quality scores are unknown, the departmental average quality score was used as the best available proxy. This is suitable because there is sufficient variety in the average research scores between departments to give moderately different averages. In the absence of bias, the higher the correlation between ChatGPT scores and the departmental average scores, the higher the underlying correlation between ChatGPT scores and the (unknown) article quality scores. The sample sizes are in Table 1.

Table 1. Sample sizes for the 34 UoAs and the combined set. Some articles are in multiple UoAs, so the total of the UoAs is smaller than that of the combined set.

| UoA | Articles |
|---|---|
| 1 - Clinical Medicine | 9556 |
| 2 - Public Health, Health Services and Primary Care | 3793 |
| 3 - Allied Health Professions, Dentistry, Nursing and Pharmacy | 9210 |
| 4 - Psychology, Psychiatry and Neuroscience | 7725 |
| 5 - Biological Sciences | 6092 |
| 6 - Agriculture, Food and Veterinary Sciences | 3046 |
| 7 - Earth Systems and Environmental Sciences | 3623 |
| 8 - Chemistry | 2620 |
| 9 - Physics | 3709 |
| 10 - Mathematical Sciences | 2978 |
| 11 - Computer Science and Informatics | 4156 |
| 12 - Engineering | 14660 |
| 13 - Architecture, Built Environment and Planning | 2285 |
| 14 - Geography and Environmental Studies | 3092 |
| 15 - Archaeology | 460 |
| 16 - Economics and Econometrics | 699 |
| 17 - Business and Management Studies | 8583 |
| 18 - Law | 1426 |
| 19 - Politics and International Studies | 1949 |
| 20 - Social Work and Social Policy | 2685 |
| 21 - Sociology | 1227 |
| 22 - Anthropology and Development Studies | 749 |
| 23 - Education | 2753 |
| 24 - Sport and Exercise Sciences, Leisure and Tourism | 2594 |
| 25 - Area Studies | 400 |
| 26 - Modern Languages and Linguistics | 733 |
| 27 - English Language and Literature | 507 |
| 28 - History | 668 |
| 29 - Classics | 55 |
| 30 - Philosophy | 393 |
| 31 - Theology and Religious Studies | 136 |
| 32 - Art and Design: History, Practice and Theory | 856 |
| 33 - Music, Drama, Dance, Performing Arts, Film and Screen Studies | 429 |
| 34 - Communication, Cultural and Media Studies, Library & Information Management | 786 |
| All | 96,800 |

### *3.2 System instructions*

All prompts used reported in this article, including those repeated from a previous study comprise both system instructions and user prompts. Whereas the system prompts define the task in general terms, telling ChatGPT how to respond, the user prompts give the specific task information. The system prompts define the task of assessing the research quality of a published journal article using the instructions published in the UK REF2021 and given to the human experts to guide research quality evaluation. These human instructions (a page and a half A4) have previously been adapted into ChatGPT system instructions, mainly by changing the first sentence to "You are an academic expert, assessing academic journal articles based on originality, significance, and rigour in alignment with international research quality standards". There are separate versions for the health and life sciences (Main Panel A, UoAs 1-6), physical sciences, maths and engineering (Main Panel B, UoAs 7-12), the social sciences (Main Panel C, UoAs 13-24), and the arts and humanities (Main Panel A, UoAs 25-34). The labels (my own) for these four Main Panel groupings are also approximate with, for example, some social science research in all four. The REF instructions define the four quality levels, explain that the task involves assessing rigour, significance, and originality, and give some examples of factors that might influence the assignment of articles to quality levels.

These system instructions were used for the current paper without change for compatibility with previous research. The full instructions are in the appendix of a previous study (Thelwall & Yaghi, 2024). The following user prompt from previous papers was not used in the current paper but was used in the comparative results reported from them alongside those from the current paper.

> Score this journal article:
> [Article title]
> Abstract
> [Article abstract]

### *3.3 ChatGPT classification scores: user prompts*

ChatGPT can be asked to estimate the relative likelihood of a variety of possible responses by reporting a table of the percentage chances of the options. Since no prior study seems to have attempted this, there is no precedent for the query format. After some experiments and asking advice from ChatGPT, the following query format was used, where the title and abstract of the article to be assessed are listed at the end.

> Given the following article, estimate the likelihood (in percentages) that it belongs to each of the four quality categories and then stop. The total should add up to 100%.
> Categories:
> 1*
> 2*
> 3*
> 4*
> Respond with a list like:
> 1*: __%
> 2*: __%
> 3*: __%
> 4*: __%

Article:
[Article title]
[Article abstract]

This apparently novel prompt format uses percentages rather than probabilities, but this seems more natural and is also an implicit request for two decimal places of accuracy for the corresponding probabilities, after dividing by 100. Unfortunately, it is not practical to test variations of this approach because of the cost of the queries. The following is an example response from the above query.

1*: 10%
2*: 20%
3*: 35%
4*: 35%

The overall score is the percentage-weighted average. In the above case it would be 1× 0.1 + 2 × 0.2 + 3 × 0.35 + 4 × 0.35=2.95.

### 3.4 ChatGPT probabilities: user prompts

Although ChatGPT can't be directly asked to report the probability of the tokens representing all different potential scores, this information can be elicited indirectly by asking it for the score only and also requesting the probability of the top 5 tokens for all tokens in its request, as follows. Here, a token represents the unit of meaning incorporated by ChatGPT into its model and may be a single character or a short string of consecutive characters. If ChatGPT is asked to return the top 5 tokens for a response, then it will not just return the response but also the five most likely tokens for the response and their probabilities.

For example, if the response is "Score 4*" then this might consist of two tokens, "Score_" and "4*" The list of top 5 alternative tokens for "Score_" might be:

Score_: 0.5;
The: 01.;
Consider: 0.001;
First: 0.001;
My: 0.0002.

The top 5 alternative tokens for 4* might then be:

4*: 0.3
3*: 0.1
**: 0.1
2*: 0.1
##: 0.001

The relevant probabilities for the main four scores can be obtained from this data by parsing the tokens until one of the scores is found as a response, then checking the alternatives for that token for valid scores and then calculating the relative probabilities of all potential scores returned allocating a zero probability to all unreturned responses.

Following the above rule, the first set of probabilities can be ignored as not relevant and the second used as the main evidence. From this, the relevant probabilities are 4*: 0.3; 3*; 0.1; 2*: 0.1. Totalling the known probabilities gives 0.3+0.1+0.1=0.5. The expected probabilities for the four possible scores are then:

4*: 0.3/0.5=0.6
3*; 0.1/0.5=0.2

2*: 0.1/0.5=0.2
1*=0.

The final score estimate would then be 4×0.6+3×0.2+2×0.2=3.4. For this approach to work, the response must be constrained to be short so that the first star rating met is the final score for the article, rather than part of a discussion of potential scores. The prompt for this is as follows.

> Score this article, giving your answer as one of 1*, 2*, 3*, or 4* then stop:
> [Article title]
> Abstract
> [Article abstract]

The maximum response length for this response is set to 5, aiming to just get the score, and the requirement for logprobs was set to true, requesting the maximum number of alternative token probabilities (5). A more complex prompt would be difficult to parse to identify where the key score token (if any) so explicitly asking for a score and prompting for its immediate answer is a strategy to ensure that a score is given reliably.

The ChatGPT API delivers its information in Json format. A section of the Json response from ChatGPT is included below to show how the log probabilities are delivered and their relationship to the response. For one query the main response was, "Score: 3*" followed by two newlines.

> {"role": "assistant", "content": "Score: 3*\n\n", "refusal": null, "annotations": []},

Following this, the Json from ChatGPT included the information below about the top five log probabilities. This shows that the tokens involved are "Score", ":", "_", "3", and "*\n\n".

- "top_logprobs": [{"token": "Score", "logprob": -0.25576457381248474, "bytes": [83, 99, 111, 114, 101]}, {"token": "**", "logprob": -1.5057646036148071, "bytes": [42, 42]}, {"token": "I", "logprob": -6.130764484405518, "bytes": [73]}, {"token": "Based", "logprob": -7.130764484405518, "bytes": [66, 97, 115, 101, 100]}, {"token": "3", "logprob": -7.380764484405518, "bytes": [51]}],

- "top_logprobs": [{"token": ":", "logprob": 0.0, "bytes": [58]}, {"token": ":**", "logprob": -17.625, "bytes": [58, 42, 42]}, {"token": ":\n\n", "logprob": -20.25, "bytes": [58, 10, 10]}, {"token": ":\n", "logprob": -20.75, "bytes": [58, 10]}, {"token": " Assessment", "logprob": -21.0, "bytes": [32, 65, 115, 115, 101, 115, 115, 109, 101, 110, 116]}],

- "top_logprobs": [{"token": " ", "logprob": -0.0024806505534797907, "bytes": [32]}, {"token": " **", "logprob": -6.002480506896973, "bytes": [32, 42, 42]}, {"token": " ***", "logprob": -12.252480506896973, "bytes": [32, 42, 42, 42]}, {"token": " *", "logprob": -16.12748146057129, "bytes": [32, 42]}, {"token": " \n\n", "logprob": -16.62748146057129, "bytes": [32, 10, 10]}],

- "top_logprobs": [{"token": "3", "logprob": -0.003229052061215043, "bytes": [51]}, {"token": "2", "logprob": -5.753229141235352, "bytes": [50]}, {"token": "4", "logprob": -9.878229141235352, "bytes": [52]}, {"token": " ", "logprob": -17.37822914123535, "bytes": [32]}, {"token": " three", "logprob": -19.62822914123535, "bytes": [32, 116, 104, 114, 101, 101]}],

- "top_logprobs": [{"token": "*\n\n", "logprob": -0.0009120595059357584, "bytes": [42, 10, 10]}, {"token": "*", "logprob": -7.000912189483643, "bytes": [42]}, {"token": "*\n", "logprob": -15.375911712646484, "bytes": [42, 10]}, {"token": "

*\n\n", "logprob": -16.625911712646484, "bytes": [32, 42, 10, 10]}, {"token": "\n\n", "logprob": -17.250911712646484, "bytes": [10, 10]}]]

In the above Json extracts, the key token is the fourth one, "3", which gives the score number. In its section, the five top alternative choices for the fourth token are:

- "3", "logprob": -0.003229052061215043,
- "2", "logprob": -5.753229141235352
- "4", "logprob": -9.878229141235352,
- " ", "logprob": -17.37822914123535,
- " three", "logprob": -19.62822914123535,

Numbers written in words were rare and were ignored. The log probabilities can be exponentiated to get probabilities and then scaled to get an estimated probability for each one.

- 3: exp(-0.003229052061215043) = 0.996776156
- 2: exp(-5.753229141235352) = 0.00317252
- 4: exp(-9.878229141235352) = 0.000051279

The scores on the right are then used as the score probabilities after dividing by their sum (0.999999954), with 0 assigned to the missing 1. The final score estimate is then the weighed sum:0.996776201×3 + 0.00317252×2+0.000051279×4=2.996878623, which is 2.997 to three decimal places.

### 3.5 Analysis

Assessing the value of the score predictions does not entail calculating their accuracy because ChatGPT seems to work on a different scale to expert evaluators: It uses a narrower range of scores (e.g., mostly clustered around 3* for some fields), and has a different average. Thus, the scores need to be rescaled or normalised to be useful. In this context, the accuracy of the scores is irrelevant and the only important measure is the extent to which they correlate with human judgement.

For RQ1 and RQ2, the results were thus correlated with the departmental quality scores rather than checking accuracy directly. Also, the most important property of the result is the rank order since the necessary rescaling may not be linear. Thus, the Spearman correlation between the ChatGPT scores and departmental average scores is the core evaluation to make.

Two separate comparisons were made, one against scores from standard prompts, as reported in previous research, and one against the scores recommended by the two new prompts. As illustrated above, the standard prompt is the same as the token probability prompt except that the first line is just "Score this journal article:" and the context window was set to 1000 tokens, allowing a longer response (often with the score given only after a discussion of the merits of the paper). The results for this were taken from a previous paper (Thelwall, 2025c). The first comparisons assess whether the new strategies improve on the state-of-the-art in terms of score prediction, and the second comparisons assess whether the score probabilities improve on extracting the single best score from the revised prompts. A positive answer to the second would suggest that the probability strategy is promising, even if it gives a negative answer to the first comparison.

For RQ3, and for both new prompting strategies, all score distributions were extracted from the results (e.g., 1*: 0%, 2*: 10%, 3*: 60%, 4*: 30%). Next, for each distribution, the number of items classified 1*, 2*, 3*, and 4* was calculated for all predictions for articles where the distribution prediction occurred. For example, suppose that one article had the above distribution twice, and three other distributions, all with the highest percentage being 4*. Then the first occurrence of the above pattern would have four other predictions for the same article (4* x 3; 3* x 1), as would the second so this single article would add the combined total (4* x 6; 3* x 2) to the record for 1*: 0%, 2*: 10%, 3*: 60%, 4*: 30%. If there were no other articles predicting this distribution then the distribution of actual scores matching the above prediction would be: 1*: 0, 2*: 0, 3*: 2/8, 4*: 6/8, or in percentages 1*: 0%, 2*: 0%, 3*: 25%, 4*: 75%, which is a sizable discrepancy from the prediction. If there were other articles with the same prediction, then they would all be combined to give a single calculation for the distribution.

The extent of the disagreement between the actual and predicted scores was assessed with the Mean Absolute Deviation (MAD). Here zero would indicate that the predicted score distribution exactly matched the observed score distribution. The MAD was calculated once for each distribution weighted by the number of occurrences of that distribution.

## 4  Results

### 4.1  RQ1 and RQ2: The performance of classification percentage requests and token probability leveraging

Recall that the classification table prompt approach involved requesting a table of percentage likelihoods for the different scores and then weighting each score by its percentage. This performed substantially worse than the previously used standard ChatGPT prompt (Figure 1, particularly the "All" bars). For the classification table approach, extracting the score with the highest percentage in the table ("Classification winners" in Figure 1) gave better results than weighting each score by its percentage ("Classification percentages" in Figure 1), but still much worse than the standard prompt. Thus, the classification table approach does not work in any sense.

In contrast to the classification tables, the probability token leveraging approach of requesting a score and then, when the first score token is reached in the response, weighting the score tokens by their probabilities ("Probabilities" in Figure 1), improves on the standard prompt in the sense of the strength of correlation with the departmental REF mean (Figure 1, particularly the "All" bars). The main response from the simple prompt ("Probability winners" in Figure 1) has a weaker correlation than the standard prompt, showing that it is the underlying probabilities rather than the altered prompt (including the shorter context window) that is beneficial.

At the level of individual UoAs, the probability-weighted scores have the highest correlation for 24 of the 34 UoAs, but these include 98% of all articles, so it is only the smaller UoAs for which the approach does not seem to work as well. For these other UoAs, classification percentages work best in five UoAs, the standard approach in four, and the classification winner in one (the smallest). In terms of broad disciplinary areas, the probability-weighted scores work best in all health and life sciences (Main Panel A), physical sciences, maths and engineering (B) and nearly all social sciences (C). For the three social sciences that are exceptions, the probability-weighted scores have the second highest correlation and are not far behind the highest correlation approach (classification percentages in all cases). It is thus the arts and humanities (D, including some social science areas), that it the probability-weighted scores perform the worst in the sense that they are the optimal strategy in only three of the ten areas.

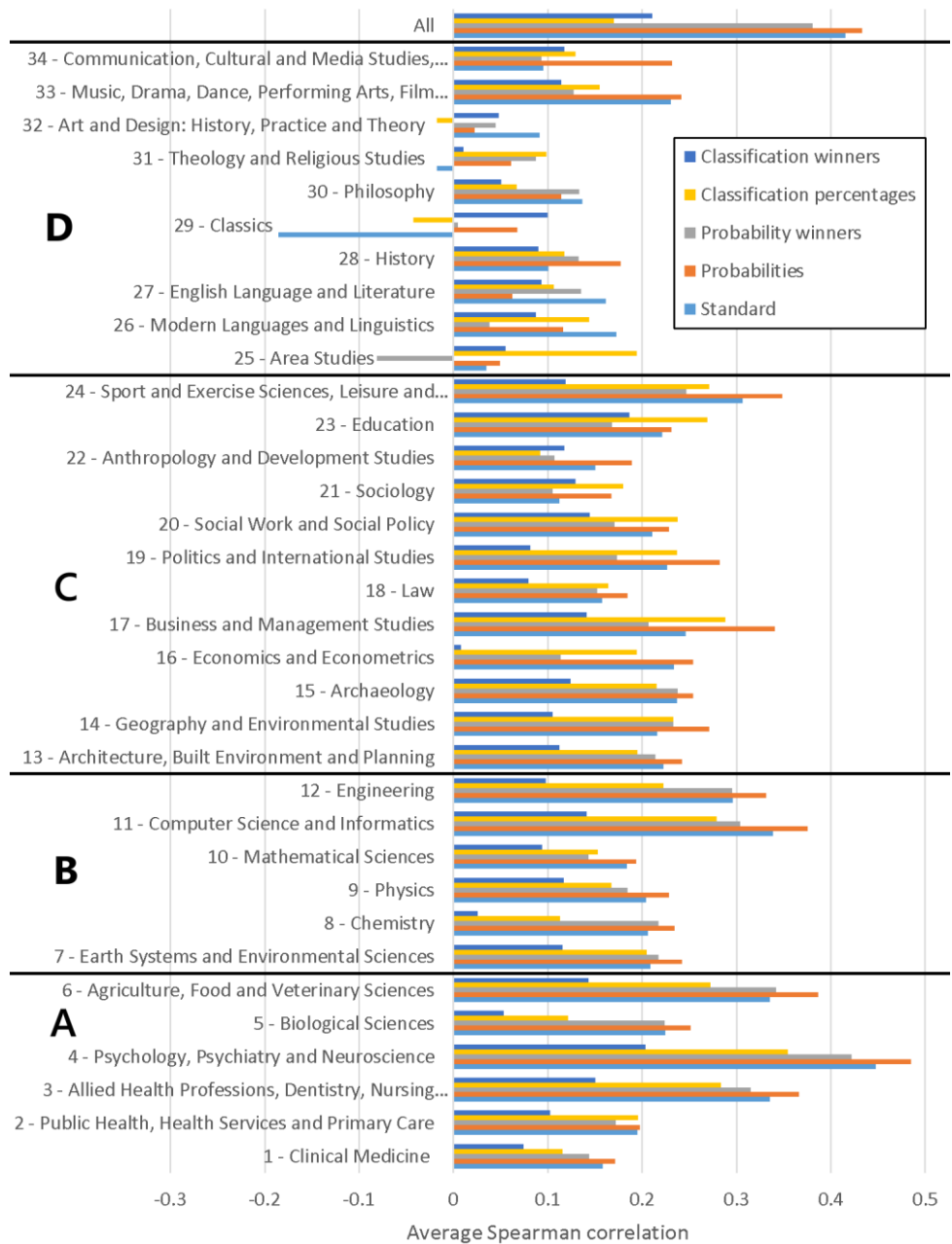

Figure 1. Average (across 2014 to 2020) Spearman correlation between departmental REF mean and ChatGPT 4o-mini scores using (a) the standard prompt, (b) the classification table prompt and taking the highest percentage score "winner" or weighting the scores by their percentages, or (c) the simple score prompt and taking the highest probability score "winner" or weighting the scores by their probability. The standard prompt values are from a previous study (Thelwall, 2025c).

## 4.2 RQ3: The internal consistencies of probabilities in classification percentage tables and score tokens

Irrespective of the extent to which the ChatGPT scores correlate with the gold standard, checking whether the predicted score profiles from both approaches match the range of scores from the same prompts assesses the internal consistency of the reported probabilities and percentages.

For each score profile prediction (classification percentages or token probabilities) the scores given by other prompts for the same article were counted (and totalled over all articles with the same prediction profile) to check whether the classification tables and token probabilities were consistent with the ChatGPT outputs. Using Mean Absolute Deviation as a simple measure, the token probability approach (A-D probabilities in Figure 2) closely - but not exactly - matches ChatGPT's outputs, whereas the classification percentage predictions (A-D prediction in Figure 2) have a much poorer match (Figure 2). This shows that the latter approach is internally inaccurate in predicting the scores given by ChatGPT. The values are in percentages and can be higher than 100% because each MAD is the sum of up to 4 prediction errors, one for each of the star levels (See Figure 3 for examples).

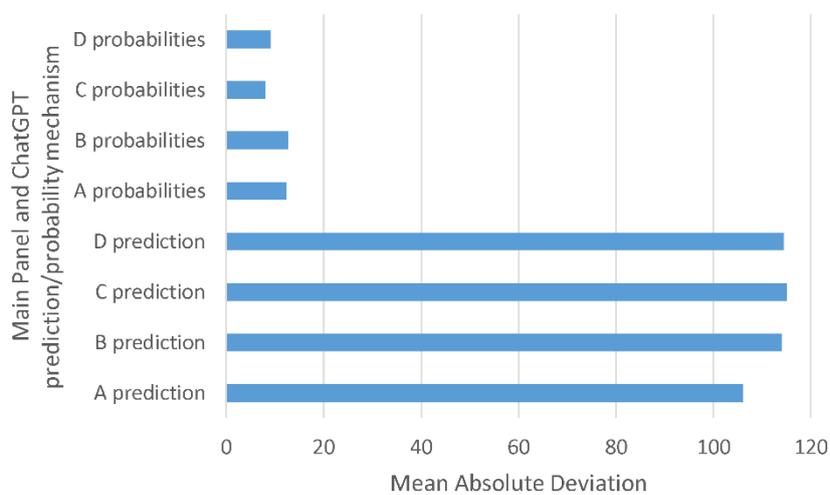

Figure 2. Mean absolute deviation (in percentage points) between the predicted (up to four) score percentages or probabilities and the actual score distribution for other predictions for the same article.

### 4.2.1 Classification percentage requests

For the classification percentages, the discrepancies between the predicted scores and actual scores for the most common predictions in each Main Panel are evident for the most common set of percentages (10-20-40-30: a 10% chance of 1*, 20% chance of 2*, 40% chance of 3* and 30% chance of 4*). Graphs for other Main Panels are very similar. In particular, the prediction is that 40% of the ChatGPT scores for an article will be 3* whereas more than 90% of the scores are 3* for these articles in each Main Panel (Figure 3).

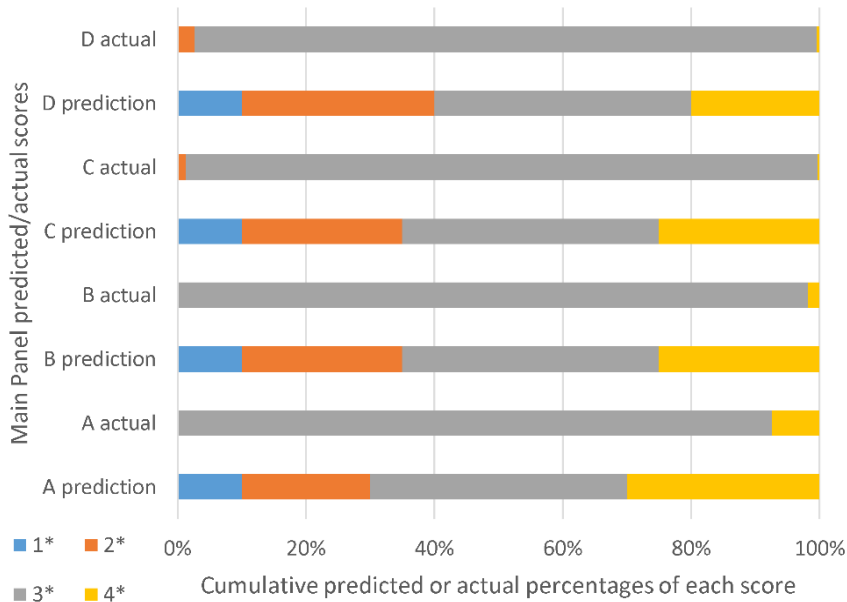

Figure 3. Predicted and actual percentages for the most common prediction percentages in each Main Panel.

The classification score percentages were usually the same for different iterations of the same article prompt but were not always identical (Table 2). ChatGPT's responses were almost always just lists of percentages in the requested format and nothing else (e.g., Table 2). The main source of variation was the presence or absence of spaces after the percentage signs. Short reviews were occasionally added after the table, however, such as, "This assessment reflects the likelihood of the article fitting into each quality category based on its originality, significance, and rigor. The work demonstrates some innovative aspects and engages with complex issues, indicating a reasonable potential for international influence and scholarly dialogue, but does not reach the highest global standards seen in 4* categories.", and "This distribution reflects the article's recognition and contribution to the understanding of a specific literary context, while indicating that it has not reached world-leading or groundbreaking status but offers significant insights, particularly in relation to Romantic literature and literary tourism."

Table 2. Responses and classification percentages for five iterations of the same query, illustrating that the estimated percentages can change for identical inputs.

| Iteration | Response | 1* perc. | 2* perc. | 3* perc. | 4* perc. | Mean score |
|---|---|---|---|---|---|---|
| 1 | 1*: 10% \n2*: 30% \n3*: 45% \n4*: 15% | 10 | 30 | 45 | 15 | 2.65 |
| 2 | 1*: 10% \n2*: 30% \n3*: 40% \n4*: 20% | 10 | 30 | 40 | 20 | 2.7 |
| 3 | 1*: 10% \n2*: 30% \n3*: 40% \n4*: 20% | 10 | 30 | 40 | 20 | 2.7 |
| 4 | 1*: 10% \n2*: 25% \n3*: 40% \n4*: 25% | 10 | 25 | 40 | 25 | 2.8 |
| 5 | 1*: 15% \n2*: 25% \n3*: 40% \n4*: 20% | 15 | 25 | 40 | 20 | 2.65 |

### 4.2.2 Token probability leveraging

For the token probabilities, there is a relatively small discrepancy between the predicted scores and actual scores for the most common predictions in each Main Panel (Figure 2). This is illustrated with common profiles for the token probability approach (Figure 4). The top profiles were a much closer match than these, so the graph illustrates the potential for discrepancies rather than typical discrepancies.

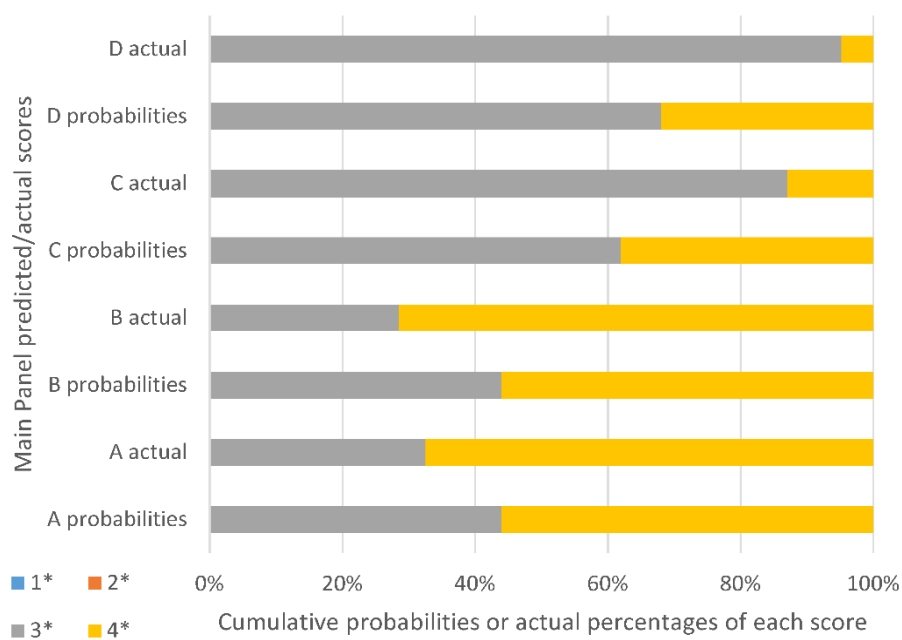

Figure 4. Token probabilities (converted to %) and actual percentages for selected common token probabilities in each Main Panel.

The token probabilities were often the same for different queries but not always. Although the biggest changes seemed to occur when the response varied, the token probabilities sometimes differed for identical responses (e.g., iterations 2 and 5 in Table 3). The responses were almost always in the format shown in Table 3, although there were rare variations, such as, "3*\n\n**Originality", "Score: **3***\n\n", and "Based on the provided abstract" (which did not yield a score).

Table 3. Responses and token probabilities for five iterations of the same query, illustrating that the underlying token probabilities can change for identical inputs.

| Iteration | Response | 1* prob. | 2* prob. | 3* prob. | 4* prob. | Mean score |
|---|---|---|---|---|---|---|
| 1 | **Score: 3 | 0 | 0 | 90 | 10 | 3.1 |
| 2 | Score: 3*\n\n | 0 | 0 | 82 | 18 | 3.18 |
| 3 | Score: 3*\n\n | 0 | 0 | 92 | 8 | 3.08 |
| 4 | Score: 3*\n\n | 0 | 0 | 82 | 18 | 3.18 |
| 5 | Score: 3*\n\n | 0 | 0 | 92 | 8 | 3.08 |

### 4.2.3 Examples of common distributions for both approaches

Some examples of the most common distributions are given here to illustrate the extent and nature of the discrepancies between the actual and predicted results. For the classification percentage requests, more than half of the prompts returned an identical class profile: 10-20-40-30 (Figure 5). Moreover, almost all score profiles include a 10% chance of a 1* score, despite this score never being allocated. This confirms that the classification percentage tables do not match ChatGPT's output well.

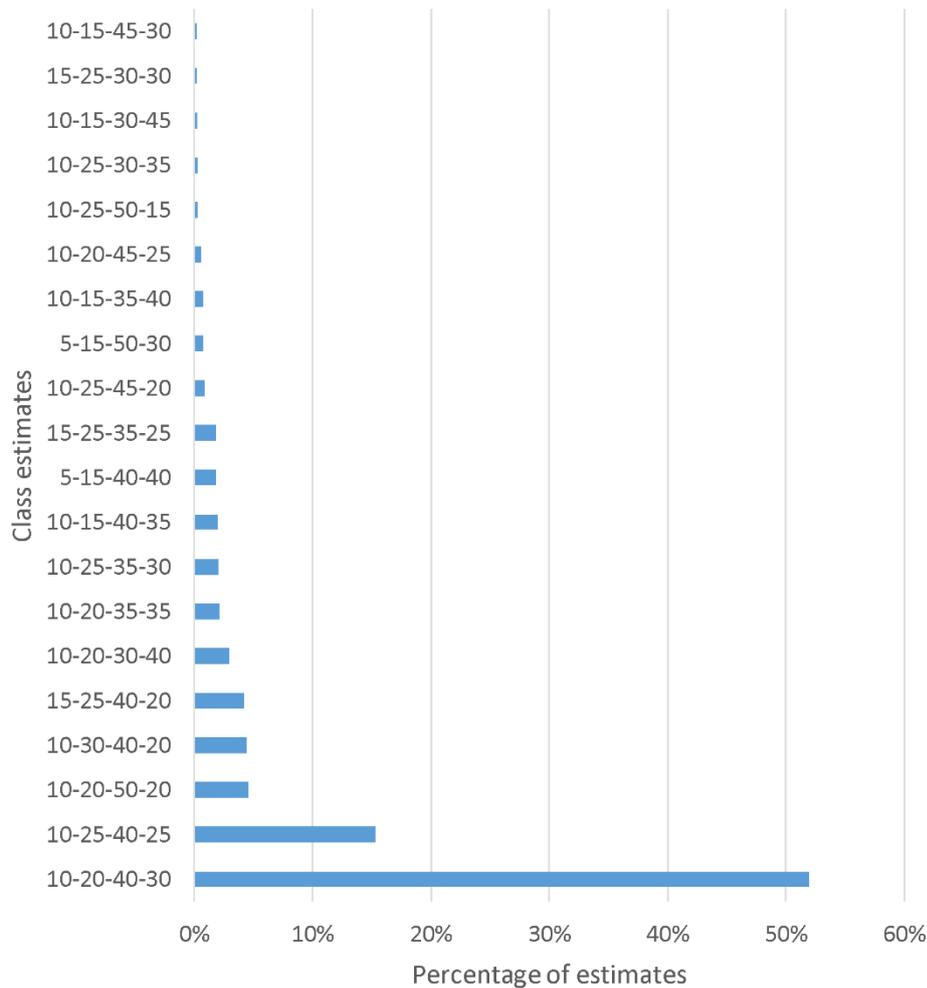

Figure 5. The 20 most common classification percentage estimates for Main Panel A (UoAs 1-6). Each number in the list is the reported percentage likelihood of the score being in the corresponding category. For example, 10-20-40-30 means 10% chance of 1*, 20% chance of 2*, 40% chance of 3* and 30% chance of 4*. Graphs for other Main Panels are very similar.

For the token probability approach, there is not a single dominant score profile, although 100% certainties about a 3* or 4* score are the most common (Figure 6).

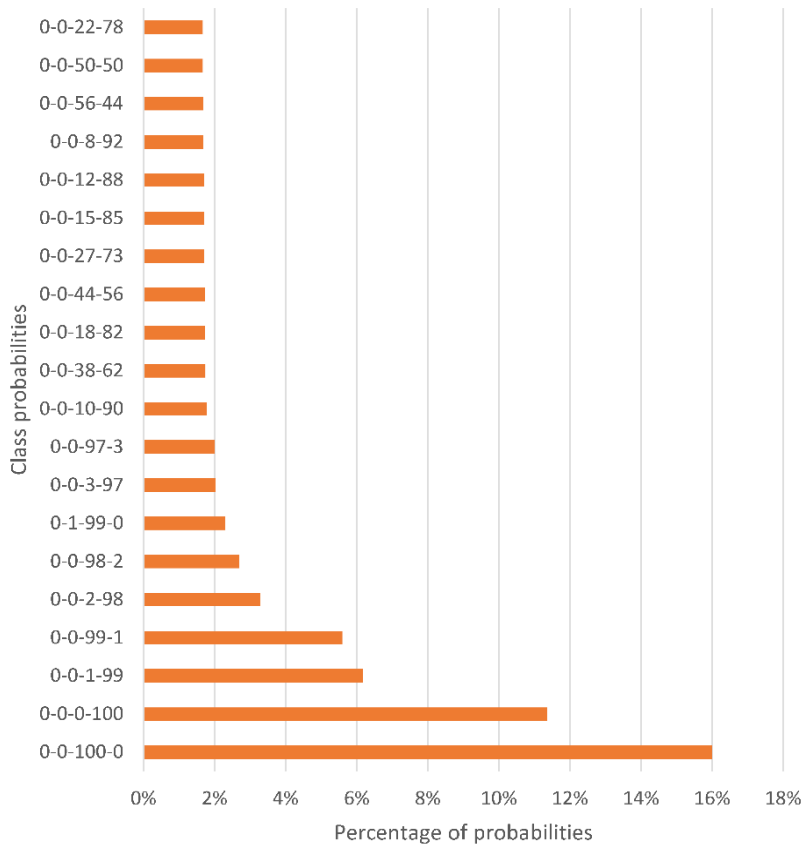

Figure 6. The 20 most common token probability estimates (converted to percentages) for Main Panel A (UoAs 1-6). Each number in the list is the calculated probability (as a percentage) the score token being in the corresponding category. For example, 0-0-99-1 means a 99% probability of 3*, and a 1% probability of 4*, according to ChatGPT's estimates. Graphs for other Main Panels are very similar.

## 5   Discussion

The results align with a previous finding that explicit (single) self-reported answer confidence percentages or probabilities given in LLM answers have less value than implicit probabilities extracted from actual and potential answer tokens (Gu et al., 2024). It extends this to a different context (sets of scores rather than single estimates) and a much larger LLM (ChatGPT 4o-mini). Whilst the current results do not rule out the possibility that an alternative prompt format would have given more useful scores, the fact that the results broadly align with previous research suggests that modifying the prompt requesting self-reported percentage confidence values is not a promising approach.

The failure of the table of self-reported confidence levels seems to stem from ChatGPT 4o favouring a particular set of percentages (10%, 20%, 40%, 30%), suggesting that it may be picking a likely pattern with little input from the specific task that it has been given. It is possible that the complexity of the request to produce a table of percentages overloads its reasoning capability, or the table request is too unusual to harness much of its prior instruction-following learning (Ouyang et al., 2022). In contrast, the request to report token probabilities is a natural part of the algorithm and therefore consumes no reasoning resources. There is no evidence in support of this intuitive explanation, however, and even some against it in the sense of an experiment showing that ChatGPT 4o being able to cope well with multiple simultaneous tasks in a single prompt (Son et al., 2024).

The simple ChatGPT prompt used for the token probability calculations was less effective than the standard prompt when the token probabilities were not calculated but just the "winning" token was used. This tends to support, on a larger scale than before (Thelwall, 2025a), that allowing ChatGPT to discuss an article before giving its score supports a better judgement. Longer reports can be thought of as a variation of the chain-of-thought approach (Wei et al., 2022) because it encourages reasoning before the final decision. This occurs despite the chain-of-thought approach reducing the accuracy of self-reported answer confidence levels (Fu et al., 2025), which is relevant only to the first strategy used here.

The results are limited by a single UK-based dataset. The gold standard is public, albeit implicit and in disconnected parts, so ChatGPT could theoretically have "cheated" by connecting the parts and then connecting these parts to the article titles and abstracts submitted to help pick a score. Mitigating against this, other studies have found correlations with private research quality data (e.g., Thelwall, 2025). Although the prompts excluded all information about each article except its title and abstract, ChatGPT may have found information about it online, such as author or journal prestige, and using it to nudge the predicted score. Moreover, any "cheating" would not give an obvious advantage for one ChatGPT strategy over another and so would not affect the conclusions.

Another limitation is that other approaches may also have been tried. Fine tuning may have given different conclusions, as might a substantially different number of iterations (e.g., 100 instead of 5). Finally, only one LLM was tested and other LLMs or ChatGPT versions may have given different patterns.

# 6   Conclusions

The results show that token probabilities reported by ChatGPT 4o-mini can be exploited to gain more useful and cheaper quality score predictions for journal articles than previous methods. Although the improvements are small and not universal, they are convincing because they occur in the UoAs accounting for 98% of the articles assessed. Conversely, asking ChatGPT to explicitly state its level of certainty for all possible answers gives results that are both weak and inconsistent, confirming that explicit likelihood requests in prompts are unhelpful.

The token probability method seems to work by leveraging ChatGPT's internal uncertainty about its answers, adjusting the main predicted score with this information. Although no other models have been tested, since ChatGPT 4o and 4o-mini represent the state of the art for the task of journal article quality score estimation, this finding also increases the relative usefulness of LLMs for research quality evaluation, by generating cheaper and more accurate article rankings. This does not solve the problem of needing to scale the ChatGPT results to give a similar average and distribution as human experts, however.

The improved accuracy of the probability approach seems to work despite the simpler prompts needed for it being less effective at predicting article scores (i.e., the lower correlation for probability winners than for standard prompts in Figure 1). Thus, it is still possible that the most accurate results would be obtained by submitting large numbers of standard prompts (e.g., 100) and averaging the scores. It would be expensive to test this, but it is a strategy that should be tried when the need for accuracy outweighs the costs. In other contexts, the results suggest that the optimal approach for obtaining a research quality indicator is to use ChatGPT five times with the probability method introduced here, averaging the five weighted average of the score predictions.

Finally, recall that practical uses of this data should consider the fact that ChatGPT scores are guesses derived from comparing article titles and abstracts with REF2021 quality evaluation guidelines rather than genuine evaluations of the articles. In other words, ChatGPT scores give a research quality indicator rather than a measure or assessment of research quality. Since these scores operate on a different scale to the REF reviewers, they are more useful to rank articles than to estimate their expert scores.

# 7  Acknowledgement

This study is funded by the Economic and Social Research Council (ESRC), UK (APP43146). For the purpose of open access, the author has applied a Creative Commons Attribution (CC BY) licence to any Author Accepted Manuscript version arising.
**Data Availability**: For legal reasons, data from Scopus cannot be made openly available. The ChatGPT software is available online https://github.com/MikeThelwall/LargeLanguageModels, as is the software to extract scores from the code for standard prompts https://github.com/MikeThelwall/Webometric_Analyst.